\documentclass{emulateapj}

\usepackage{natbib}
\slugcomment{Accepted for Pub in ApJ, October 12, 2009}

\begin{document}
\renewcommand{\arraystretch}{0.9}

\title{\textit{Fermi} LAT detection of pulsed $\gamma$-rays from the Vela-like pulsars PSR J1048$-$5832 and PSR J2229+6114}

\author{
A.~A.~Abdo\altaffilmark{1,2}, 
M.~Ackermann\altaffilmark{3}, 
M.~Ajello\altaffilmark{3}, 
W.~B.~Atwood\altaffilmark{4}, 
M.~Axelsson\altaffilmark{5,6}, 
L.~Baldini\altaffilmark{7}, 
J.~Ballet\altaffilmark{8}, 
G.~Barbiellini\altaffilmark{9,10}, 
M.~G.~Baring\altaffilmark{11}, 
D.~Bastieri\altaffilmark{12,13}
% ...
}
\altaffiltext{1}{Space Science Division, Naval Research Laboratory, Washington, DC 20375, USA}
\altaffiltext{2}{National Research Council Research Associate, National Academy of Sciences, Washington, DC 20001, USA}
\altaffiltext{3}{W. W. Hansen Experimental Physics Laboratory, Kavli Institute for Particle Astrophysics and Cosmology, Department of Physics and SLAC National Accelerator Laboratory, Stanford University, Stanford, CA 94305, USA}
\altaffiltext{4}{Santa Cruz Institute for Particle Physics, Department of Physics and Department of Astronomy and Astrophysics, University of California at Santa Cruz, Santa Cruz, CA 95064, USA}
\altaffiltext{5}{Department of Astronomy, Stockholm University, SE-106 91 Stockholm, Sweden}
\altaffiltext{6}{The Oskar Klein Centre for Cosmoparticle Physics, AlbaNova, SE-106 91 Stockholm, Sweden}
\altaffiltext{7}{Istituto Nazionale di Fisica Nucleare, Sezione di Pisa, I-56127 Pisa, Italy}
\altaffiltext{8}{Laboratoire AIM, CEA-IRFU/CNRS/Universit\'e Paris Diderot, Service d'Astrophysique, CEA Saclay, 91191 Gif sur Yvette, France}
\altaffiltext{9}{Istituto Nazionale di Fisica Nucleare, Sezione di Trieste, I-34127 Trieste, Italy}
\altaffiltext{10}{Dipartimento di Fisica, Universit\`a di Trieste, I-34127 Trieste, Italy}
\altaffiltext{11}{Rice University, Department of Physics and Astronomy, MS-108, P. O. Box 1892, Houston, TX 77251, USA}
\altaffiltext{12}{Istituto Nazionale di Fisica Nucleare, Sezione di Padova, I-35131 Padova, Italy}
\altaffiltext{13}{Dipartimento di Fisica ``G. Galilei", Universit\`a di Padova, I-35131 Padova, Italy}
% ...

\begin{abstract}
We report the detection of $\gamma$-ray pulsations ($\ge 0.1$ GeV) from PSR~J2229+6114 and PSR~J1048$-$5832, the latter having been detected as a low-significance pulsar by EGRET. Data in the $\gamma$-ray band were acquired by the Large Area Telescope aboard the \textit{Fermi Gamma-ray Space Telescope}, while the radio rotational ephemerides used to fold the $\gamma$-ray light curves were obtained using the Green Bank Telescope, the Lovell telescope at Jodrell Bank, and the Parkes telescope. The two young radio pulsars, located within the error circles of the previously unidentified EGRET sources 3EG~J1048$-$5840 and 3EG~J2227+6122, present spin-down characteristics similar to the Vela pulsar. PSR~J1048$-$5832 shows two sharp peaks at phases $0.15 \pm 0.01$ and $0.57 \pm 0.01$ relative to the radio pulse confirming the EGRET light curve, while PSR~J2229+6114 presents a very broad peak at phase $0.49 \pm 0.01$. The $\gamma$-ray spectra above 0.1 GeV of both pulsars are fit with power laws having exponential cutoffs near 3 GeV, leading to integral photon fluxes of $(2.19 \pm 0.22 \pm 0.32) \times 10^{-7}$\,cm$^{-2}$\,s$^{-1}$  for PSR~J1048$-$5832 and $(3.77 \pm 0.22 \pm 0.44) \times 10^{-7}$\,cm$^{-2}$\,s$^{-1}$ for PSR~J2229+6114. The first uncertainty is statistical and the second is systematic. PSR~J1048$-$5832 is one of two LAT sources which were entangled together as 3EG~J1048$-$5840. These detections add to the growing number of young $\gamma$-ray pulsars that make up the dominant population of GeV $\gamma$-ray sources in the Galactic plane.
   
\end{abstract}

\keywords{gamma rays: observations --- pulsars: general --- pulsars: individual (PSR J1048$-$5832, PSR J2229+6114)}

\section{Introduction}

The nature of unidentified high-energy $\gamma$-ray sources in the Galaxy was one of the major unanswered questions at the end of the EGRET era. The third EGRET catalog contained 170 unidentified sources, 74 of which were at Galactic latitude $|b|<10$\degr \citep{hartman99,bhatta03}. Rotation-powered pulsars are believed to dominate the Galactic $\gamma$-ray source population (e.g. \citet{yadiga95}), but their visibility is linked to their beam patterns. Soon after launch, the {\it Fermi Gamma-ray Space Telescope} began to unveil many 3EG sources, discovering the radio-quiet pulsar in the CTA1 supernova remnant associated with 3EG~J0010+7309 \citep{abdo08}, detecting the radio-loud pulsar PSR~J2021+3651 (associated with 3EG~J2021+3716) \citep{abdo09b}, seen independently by {\em AGILE} \citep{halpern08}, as well as the radio pulsar PSR~J1028$-$5819, associated with 3EG~J1027$-$5817 \citep{abdo09c} and new populations of radio-quiet $\gamma$-ray pulsars, detectable using blind search techniques \citep{abdo09f}. For the pulsars detected in $\gamma$ rays, the bulk of the electromagnetic power output is in high energies. The $\gamma$-ray emission is thus crucial for understanding the emission mechanism which converts the rotational energy of the neutron star into electromagnetic radiation. The discovery of many new $\gamma$-ray pulsars will provide strong constraints on the location of the $\gamma$-ray emitting regions, whether above the polar caps \citep{daugherty96}, or far from the neutron star in the so-called ``outer gaps" \citep{romani96}, or in the intermediate regions like the ``slot gap" \citep{muslimov04} having ``two-pole caustic" geometry \citep{dyks03}. In this paper we report the \textit{Fermi} detection of the two pulsars PSR~J1048$-$5832 and PSR~J2229+6114, which have spin characteristics similar to other young pulsars typified by the Vela pulsar. \citet{kramer03} provide a discussion of ``Vela-like'' pulsars. \\

PSR~J1048$-$5832 (B1046$-$58) is located in the Carina region at low Galactic latitude ($l=287.42$\degr, $b=0.58$\degr). It was discovered during a 1.4 GHz Parkes survey of the Galactic plane and has a period  $P~\sim$123.7~ms \citep{johnston92}. It has a spin-down luminosity $\dot{E}=4 \pi^{2} I (\dot{P}/P^{3})$ of $2\times 10^{36}$\,erg\,s$^{-1}$, for a moment of inertia $I$ of $10^{45}$\,g\,cm$^{2}$, a surface dipole magnetic field strength of $3.5 \times 10^{12}$\,G, and a characteristic age $\tau_{c} = P/2\dot{P}$ of 20\,kyr. High-resolution observations of the coincident $ASCA$ source by the \textit{Chandra X-ray Observatory} and {\it XMM-Newton\/} revealed an asymmetric pulsar wind nebula (PWN) of $\sim6''\times11''$, surrounding a point source coincident with the pulsar but so far no X-ray pulsations were detected \citep{gonzalez06}. PSR~J1048$-$5832 has been proposed as a counterpart of the steady EGRET source 3EG~J1048$-$5840 \citep{fierro95,pivovaroff00,nolan03}, as suggested by positional coincidence, spectral, and energetic properties. Detailed analysis of the EGRET data found possible $\gamma$-ray pulsations at $E>$400 MeV, a double-peaked light curve with $\sim 0.4$ peak phase separation (Fig.\ref{fig:J1048-5832_LC_multi_energy}, \citealt{kaspi00}). An HI distance determination for the pulsar yields between 2.5 and 6.6 kpc \citep{johnston96}. The NE2001 model \citep{cordes02} assigns a distance of 2.7~kpc based partly on the HI distance determination. For this paper we adopt 3~kpc as the distance to the pulsar.\\

PSR J2229+6114 is located at ($l$,$b$) = (106.6\degr,2.9\degr) within the error box of the EGRET source 3EG~J2227+6122 \citep{hartman99}. Detected as a compact X-ray source by \textit{ROSAT} and \textit{ASCA} observations of the EGRET error box, it was later discovered to be a radio and X-ray pulsar with a period of $P$ = 51.6 ms \citep{halpern01b}. The radio pulse profile shows a single sharp peak, while the X-ray light curve at 0.8\ --\ 10 keV consists of two peaks, separated by $\Delta\phi = 0.5$. {\em AGILE} recently reported the discovery of $\gamma$-ray pulsations above 100 MeV \citep{pellizzoni09}. The pulsar is as young as the Vela pulsar (characteristic age $\tau_{c} = 10$\,kyr), as energetic ($\dot{E} = 2.2 \times 10^{37}$\,erg\,s$^{-1}$), and is evidently the energy source of the ``Boomerang'' arc-shaped PWN G106.65+2.96, suggested to be part of the supernova remnant (SNR) G106.3+2.7 discovered by \citet{joncas90}. Recently, the PWN has been detected at TeV energies by MILAGRO \citep{abdo09g}. Studies of the radial velocities of both neutral hydrogen and molecular material place the system at $\sim$ 800 pc \citep{kothes01}, while \citet{halpern01a} suggest a distance of 3 kpc estimated from its X-ray absorption. The pulsar DM, used in conjunction with the NE2001 model, yields a distance of 7.5 kpc\,, significantly above all other estimates. For this paper we again adopt a distance of 3\,kpc.

\section{Observations}

\subsection{Gamma-Ray Observations}

The Large Area Telescope (LAT) aboard \textit{Fermi} is an electron-positron pair conversion telescope sensitive to $\gamma$-rays of energies 0.02 to $>$300 GeV. The LAT is made of a high-resolution silicon microstrip tracker, a CsI hodoscopic electromagnetic calorimeter, and segmented plastic scintillators to reject the background of charged particles \citep{atwood09}. Compared with its predecessor EGRET, the LAT has a larger effective area ($\sim$8000 cm$^{2}$ on-axis for E $>$ 1 GeV), improved angular resolution ($\theta_{68} \sim$ 0.5${^\circ}$ at 1 GeV for events in the front section of the tracker), and a higher sensitivity ($\sim 3 \times 10^{-9}$\,cm$^{-2}$\,s$^{-1}$)\footnote{This value refers to a steady source after one-year sky survey, assuming a high-latitude diffuse flux of 1.5 $\times$ 10$^{-5}$ cm$^{-2}$ s$^{-1}$ sr$^{-1}$ ( $>$0.1 GeV and a photon spectral index of -2.1 with no spectral cutoff).}. The large field of view ($\sim$2.4 sr) allows the LAT to observe the full sky in survey mode every 3 hours. The LAT timing is derived from a GPS clock on the spacecraft and $\gamma$-rays are hardware time-stamped to an accuracy significantly better than 1 $\mu$s \citep{abdo09h}. The LAT software tools for pulsars have been shown to be accurate to a few $\mu$s \citep{smith08}.

In this paper, the data used for the spectral analysis were collected during \textit{Fermi}'s first-year all-sky survey, beginning 2008 August 4. For the timing analysis, we added data collected during the commissioning phase observations from 2008 June 30 to 2008 August 3, that included pointed observations of the Vela pulsar and other targets. The data for PSR~J1048$-$5832 end 2009 April 10, while those for PSR~J2229+6114 end 2009 March 23. Only $\gamma$ rays in the ``Diffuse" class events (the tightest background rejection) were selected. In addition, we excluded those coming from zenith angles $>$~105\degr, where $\gamma$-rays resulting from cosmic-ray interactions in the Earth's atmosphere produce an excessive background contamination.

\subsection{Radio Observations}

\subsubsection{PSR~J1048$-$5832}

PSR~J1048--5832 is observed in the radio band at approximately monthly intervals since early 2007 at the 64-m Parkes radio telescope in Australia \citep{manchester08}. A typical observation is of 2~min duration at a frequency of 1.4\,GHz with occasional additional observations at 0.7 and 3.1\,GHz. Full details of the observing and data analysis can be found in \citet{welte09a}. The pulsar is known to have glitched in the past \citep{wang00} and indeed glitched just prior to the launch of Fermi \citep{welte09a}. Like many high $\dot{E}$ pulsars, PSR~J1048$-$5832 is highly polarized in the radio band over a wide frequency range \citep{kara05,johnston06}.\\
The timing solution uses 20 pulse times of arrival (TOAs) and was derived using the pulsar timing software TEMPO2 \citep{hobbs06}. The timing solution fits for the pulsar's spin frequency and frequency derivative and also whitens the data using the so-called fitwaves algorithm within TEMPO2. The resulting rms of 0.287~ms is a substantial improvement on the pre-whitened solution. Note that the extra fitwaves parameters are not supported by the standard \textit{Fermi} software tools but were included in this analysis. Measurements of the dispersion delay across the 1369\,MHz band with a bandwidth of 256 MHz lead to a dispersion measure (DM) of 128.822 $\pm$ 0.008\,pc\,cm$^{-3}$, with no indication that the DM is varying over time. This DM is used to correct the radio TOAs to infinite frequency for phasing the $\gamma$-ray photon TOAs.

\subsubsection{PSR~J2229+6114}

PSR~J2229+6114 is being observed at the NRAO Green Bank Telescope (GBT) \citep{kaplan05} and the Lovell telescope at Jodrell Bank \citep{hobbs04}. The rotational ephemeris used here to fold $\gamma$-ray photons is based on TOAs obtained from both telescopes between 2008 June 17 and 2009 March 23.  There are 25 such TOAs from GBT with average uncertainty of 0.2\,ms, each from a 5 minute observation mainly at a central frequency of 2.0\,GHz. The 44 Jodrell Bank TOAs have an average uncertainty of 0.3\,ms, obtained from individual 30 minute observations at 1.4\,GHz.  In our timing fits with TEMPO\footnote{http://www.atnf.csiro.au/research/pulsar/tempo.}, we fixed the position at that known from X-ray observations \citep{halpern01b}. PSR~J2229+6114 shows some timing noise over the 9 month interval, requiring a fit to rotation frequency $\nu = 1/P$ and its first two derivatives. In addition, on MJD $54782.6\pm0.5$ a small glitch occurred, with fractional frequency step of $\Delta \nu/\nu = (4.08\pm0.06)\times10^{-9}$ and $\Delta \dot \nu/{\dot \nu} = (2.0\pm0.4)\times10^{-4}$.  The post-fit rms is 0.24\,ms.  We have measured the DM with separate sets of GBT observations at three widely spaced frequencies, obtaining $\mbox{DM} = (204.97\pm0.02)$\,pc\,cm$^{-3}$. This is used to correct 2\,GHz arrival times to infinite frequency, for comparison with the $\gamma$-ray profile.

\section{Analysis and Results}

The events were analyzed using the standard software package \textit{Science Tools}\footnote{http://fermi.gsfc.nasa.gov/ssc/data/analysis/scitools/overview.html} (ST) for the \textit{Fermi} LAT data analysis and TEMPO2. The timing parameters used in this work will be made available on the servers of the \textit{Fermi} Science Support Center\footnote{http:$//$fermi.gsfc.nasa.gov/ssc/data/access/lat/ephems/}. 


\subsection{Pulse profiles of PSR~J1048$-$5832}

For the timing analysis of PSR J1048$-$5832, we selected $\gamma$-rays with energy $>$0.1 GeV within a radius of 1\fdg 0 around the radio pulsar position. Then, we applied an energy-dependent angular radius cut $\theta_{68} \leqslant$ $0.8 \times E_{\rm GeV}^{-0.75}$\,degrees, keeping all the photons included in a radius of 0\fdg 35. This selection approximates the LAT point spread function (PSF) and maximizes the signal-to-noise ratio over a broad energy range. We corrected photon arrival times to the Solar System barycenter using the JPL~DE405 Solar System ephemeris \citep{standish98}. The event times were then folded with the radio period using the Parkes ephemeris. The bin-independent \textit{H}-test \citep{dejager89} results in a probability $\le 4 \times 10^{-8}$ that the modulation would have occurred by chance. This value is more than 4 orders of magnitude smaller than the previous EGRET results \citep{kaspi00}, establishing firmly this source as a $\gamma$-ray pulsar. 

\begin{figure*}[]
\begin{center}
\epsscale{1.0}
\plotone{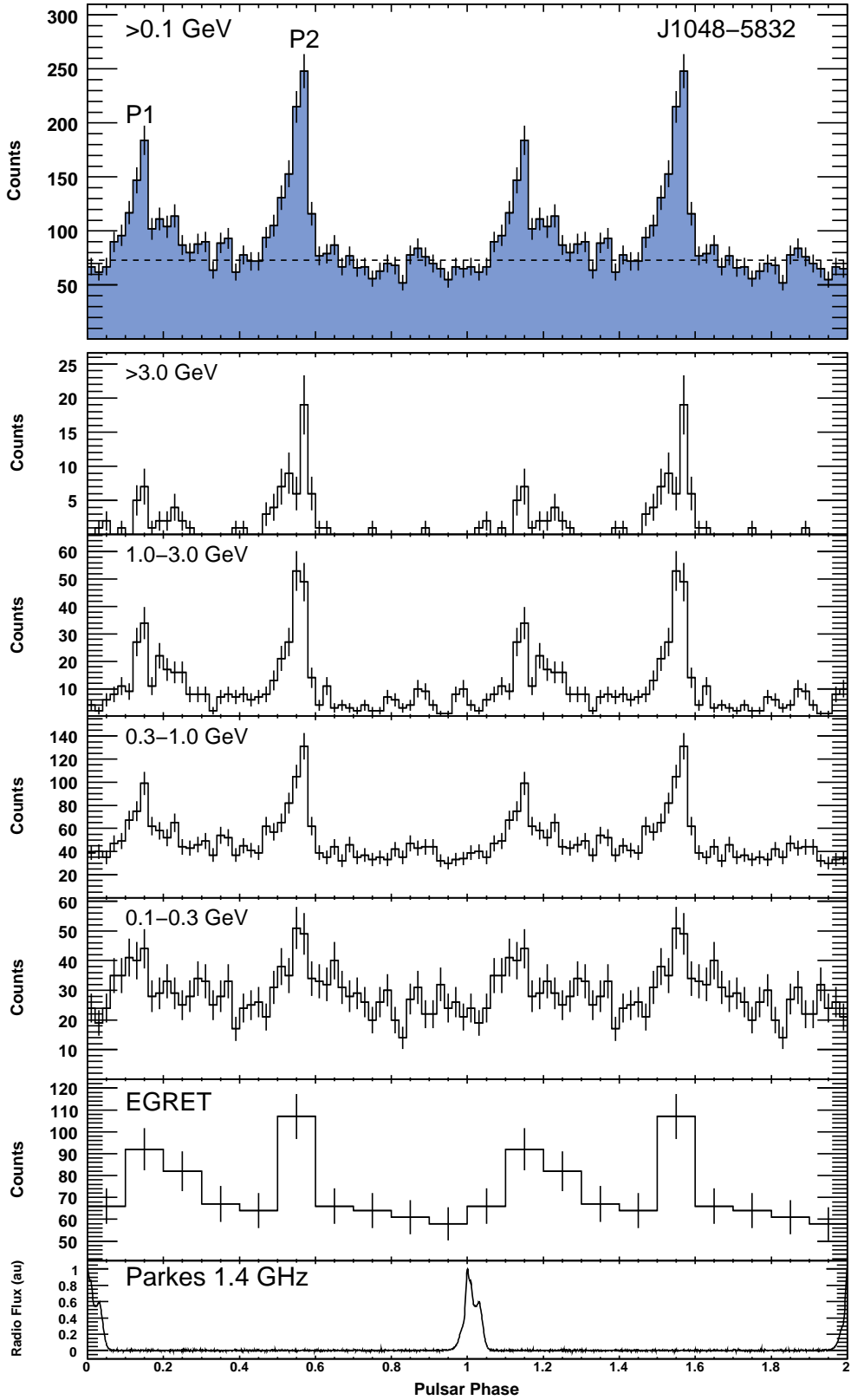}
\caption{
\textit{Top panel:}~Light curve of PSR~J1048$-$5832 above 0.1 GeV, shown over two pulse periods with 50 bins per period ($P=123.7$\,ms). The dashed line shows the background level, as estimated from an annulus surrounding the pulsar position during the off-pulse phase (73 counts/bin).  \textit{Four following panels:}~Energy-dependent phase histograms for PSR~J1048$-$5832 in the four indicated energy ranges, each displayed with 50 bins per pulse period. \textit{Second panel from bottom:}~Folded EGRET light curve for energies above 400\,MeV \citep{kaspi00}. \textit{Bottom panel:}~Radio pulse profile from the Parkes Telescope at a center frequency of 1.4\,GHz with 1024 phase bins. }
\label{fig:J1048-5832_LC_multi_energy}
\end{center}
\end{figure*}

Figure~\ref{fig:J1048-5832_LC_multi_energy} (top panel) shows the resulting 50-bin $\gamma$-ray phase histogram above 0.1 GeV. The two peaks appear asymmetric, with a slow rise and a fast fall. We fit each peak with two half-Lorentzian functions, i.e. with different widths for the leading and trailing sides. P1 ($\phi = 0.05 - 0.17$) is sharper with a full width at half maximum (FWHM) of $0.06 \pm 0.01$, while P2 ($\phi = 0.45 - 0.65$) is a bit broader with an FWHM of $0.10\pm0.02$. The 1.4 GHz radio profile is shown at the bottom panel for comparison. The $\gamma$-ray light curve consists of two peaks, with P1 at phase $0.15 \pm 0.01 \pm 0.0001$ and P2 at phase $0.57 \pm 0.01 \pm 0.0001$, leading to a phase separation $\Delta\phi$ of $0.42 \pm 0.01 \pm 0.0001$. The errors are respectively the $\gamma$-ray fit uncertainty and that caused by the DM uncertainty. These results are in agreement with those found by \citet{kaspi00}. We observe structure related to a shoulder or a ``bridge'' region between 0.17 and 0.30 in phase, which trails the first $\gamma$-ray peak. Note that the profile is very similar to the Vela light curve, which consists of both two peaks separated by 0.43 in phase and a bridge region \citep {abdo09a}. We also defined the ``off-pulse region'' as the pulse minimum ($\phi = 0.7 - 1.05$). To check this assumption, we estimated the background represented by the dashed line (73 counts/bin) from a ring with 1\degr $< \theta < $2\degr~surrounding the source. Nearby sources are removed, and we normalized to the same phase space as our selection. The result is in good agreement with the off-pulse region. As a consequence, the total number of pulse photons from the pulsar is estimated at $933 \pm 93$, with a background contribution of $3654 \pm 60$ events. 

Figure~\ref{fig:J1048-5832_LC_multi_energy} (middle panels) shows the 50-bin $\gamma$-ray phase histograms of PSR J1048$-$5832 in four energy bands (0.1\ --\ 0.3 GeV, 0.3\ --\ 1\,GeV, 1\ --\ 3\,GeV, $>$3\,GeV). Evolution in the light curve shape with energy is visible, although more data are needed to constrain the peak widths. Notably, below 0.3\,GeV the first peak seems wider than at high energies. This feature could be explained by the contamination of a LAT source (0FGL 1045.6$-$5937) less than 1$^{\circ}$ from the pulsar. We also looked for an evolution of the ratio P1/P2 with energy, as seen by EGRET for Vela, Crab, Geminga, PSR~B1951+32 \citep{thompson04} and seen by the \textit{Fermi} LAT for Vela \citep{abdo09a} and PSR~J0205+6449 \citep{abdo09e}. In each energy band, we calculated the peak height with respect to the background level as estimated from the off pulse interval. The ratio shows no variation with a confidence level of $\sim$69$\%$. This is not a particularly stringent bound, and we expect to accumulate much more data to eventually detect a possible variation of P1/P2 with energy. Finally, note that between 1 and 3 GeV, the trailing shoulder P1 ($\phi$ = 0.17\ --\ 0.33) is significant and there is still evidence of it for E $>$ 3 GeV, whereas above 3 GeV the two peaks are still observed with the highest energy photon detected in P1 at 19\,GeV. 

\subsection{Pulse profiles of PSR~J2229+6114}
For the timing analysis of PSR~J2229+6114, we used the same energy-dependent selection criteria as for PSR J1048$-$5832. $\gamma$-ray TOAs were folded with the radio period using the GBT and Lovell telescope ephemeris. 

\begin{figure*}[]
\begin{center}
\epsscale{1.0}
\plotone{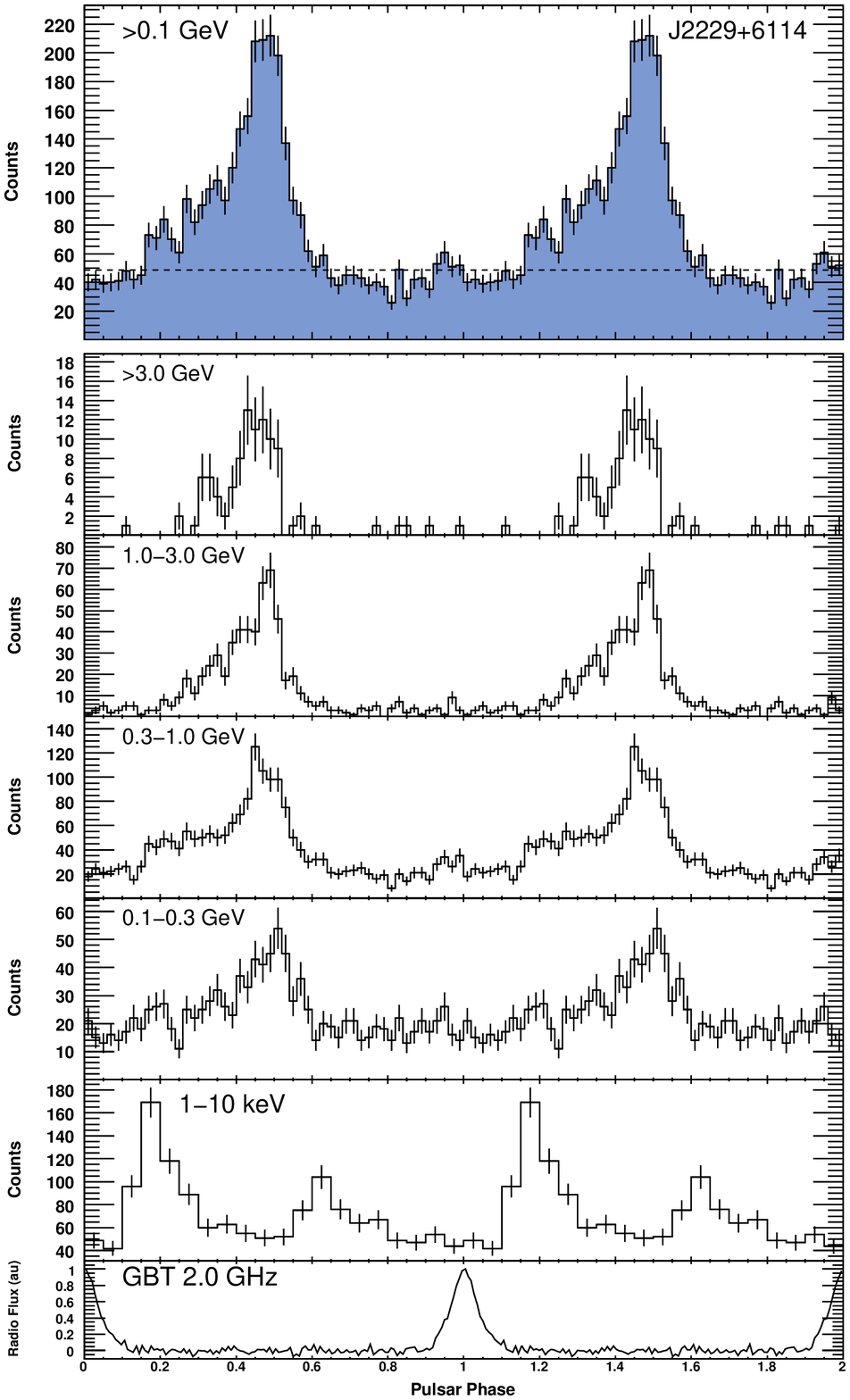}
\caption{
\textit{Top panel:}~Light curve of PSR~J2229+6114 above 0.1 GeV, shown over two pulse periods with 50 bins per period ($P=51.6$\,ms). The dashed line shows the background level, as estimated from an annulus surrounding the pulsar position during the off-pulse phase (48 counts/bin). \textit{Four following panels:}~Energy dependent phase histograms for PSR~J2229+6114 in the four indicated energy ranges, each displayed with 50 bins per pulse period. \textit{Second panel from bottom:}~Light curve in the 1--10\,keV band from the {\em XMM\/} pn CCD in small window mode.  The instrumental time resolution is 5.7\,ms (2.2 phase bins). Phase alignment with respect to the radio pulse, as described in the text, is accurate to $\approx 0.4$ phase bins of this 20-bin light curve. \textit{Bottom panel:}~Radio pulse profile from Green Bank Telescope at a center frequency of 2\,GHz with 128 phase bins.} \label{fig:J2229+6114_LC_multi_energy}
\end{center}
\end{figure*}

Figure~\ref{fig:J2229+6114_LC_multi_energy} (top panel) shows the resulting 50-bin histogram of folded counts above 0.1 GeV compared with the phase-aligned radio pulse profile (bottom panel). The profile shows a single, asymmetric peak ($\phi = 0.15 - 0.65$), that has been fit by a two half-Lorentzian function. The fit places the peak at $0.49 \pm 0.01 \pm 0.001$ with FWHM of $0.23 \pm 0.03$. We estimated the background level from a 1\ --\ 2$^{\circ}$ ring around the pulsar during the 0.65\ --\ 0.15 pulse minimum. This is represented by the dashed line (48 counts/bin). The result is consistent with the off-pulse region. As a consequence, the total of pulsed photons from the pulsar is estimated at $1365 \pm 79$, with a background contribution of $2431 \pm 49$ events.

To examine the energy-dependent trend of the $\gamma$-ray pulse profile (Figure~\ref{fig:J2229+6114_LC_multi_energy}, middle panels), phase histograms are plotted over 4 energy intervals ($0.1-0.3$\,GeV, $0.3-1$\,GeV, $1-3$\,GeV and $>$3 GeV), showing a possible drift of the peak. Between 0.1\ --\ 0.3 GeV, the peak is offset from the radio pulse by $0.51 \pm 0.02$ according to a two half-Lorentzian fit, while the offsets for 0.3\ --\ 1\,GeV, 1\ --\ 3\,GeV and $>$ 3 GeV are $0.48 \pm 0.01$, $0.49 \pm 0.01$ and $0.45 \pm 0.01$, respectively. The peak positions are compatible between 0.1\ -- 3 GeV, and a slight mis-alignment appears above 3\,GeV. However, the data cannot constrain strongly the pulse phase dependence, even if the mis-alignment between the X-ray and $\gamma$-ray profile suggests a spectral energy dependence of the high energy light curve. Finally, we note that the highest energy photon is 10.8\,GeV at phase $\sim$ 0.30.

In Figure~\ref{fig:J2229+6114_LC_multi_energy} we also show the 1--10\,keV X-ray profile of PSR~J2229+6114 obtained from an {\em XMM\/} observation on 2002 June 15 (MJD~52440) with effective exposure time of 20\,ks. The data were folded using a contemporaneous ephemeris based on Jodrell Bank and GBT observations. The highest peak of the X-ray profile lags the radio pulse by $\phi$ = $0.17\pm0.02$. Analysis of an {\em RXTE\/} observation from MJD~52250 yields a consistent radio--X-ray offset. There is no energy dependence of the X-ray pulse shape within the 1--10\,keV range, while the sharpness of the peaks as well as their spectral shape indicates that the emission is predominantly non-thermal.


\subsection{Phase-averaged spectra and flux}


In order to obtain the phase-averaged spectrum for PSR~J1048$-$5832 and PSR~J2229+6114, a maximum likelihood spectral analysis \citep{mattox96} was performed, using the LAT tool `gtlike'$^{3}$. We used the ``Pass 6 v3'' instrument response functions (IRFs), which are a post-launch update to address $\gamma$-ray detection inefficiencies that are correlated with trigger rate. The systematic errors on the effective area are $\le$ 5\% near 1 GeV, 10\% below 0.1 GeV and 20\% over 10 GeV. 


\subsubsection{Spectrum of PSR~J1048$-$5832}

\begin{figure*}
\begin{center}
\epsscale{1.}
\plotone{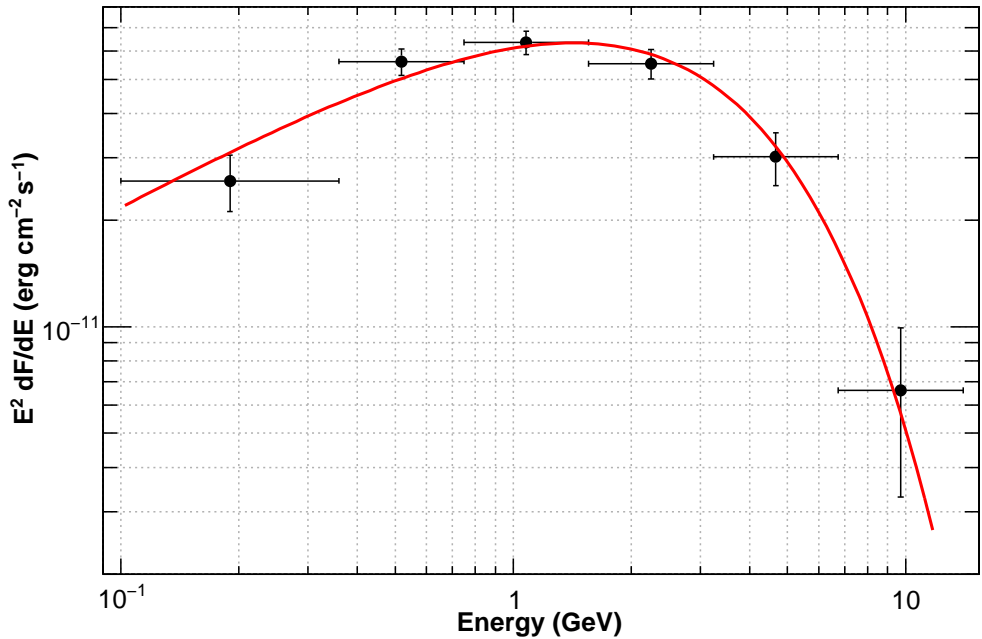}
\caption{Spectral energy distribution for PSR~J1048--5832 as fit by ``gtlike''assuming a power-law with an exponential cutoff (solid line). The spectral points were obtained for 6 logarithmically-spaced energy bins and performing spectral analysis in each interval, assuming a power law shape for the source. The errors bars are statistical only.}
\label{fig:sed_j1048}
\end{center}
\end{figure*}

A circular region around 15$^{\circ}$ from the source was modeled including nearby LAT sources around the source position. The Galactic diffuse background was accounted for by using maps based on the GALPROP model called 54\_59Xvarh7S \citep{strong04a,strong04b}. We modeled the isotropic background through a tabulated spectrum derived from a fit to LAT data at high galactic latitude, freezing the Galactic model and including the detected point sources within a radius of 15\degr. We fit the spectrum of PSR~J1048$-$5832 with a power law with an exponential cutoff between 0.1 and 0.7 in phase, that can be described by the equation:
\begin{eqnarray}\label{eq:spectrum}
\frac{dF}{dE} = N_{0} \ E^{-\Gamma} e^{-E/E_{c}}
\label{eqn_diff_spec}
\ {\rm cm^{-2} s^{-1} GeV^{-1}}
\end{eqnarray}
with $E$ in GeV, the term $N_{0} = (5.9 \pm 0.3 \pm 0.1) \times 10^{-8}$\,cm$^{-2}$\,s$^{-1}$\,GeV$^{-1}$, a spectral index $\Gamma$ = $1.38 \pm 0.06 \pm 0.12$ and a cut-off energy $E_{c} = 2.3^{+0.3}_{-0.4} \pm 0.3$\,GeV. The first uncertainty is statistical, while the second is systematic. From this result, we obtained an integral photon flux in the range 0.1-100 GeV of $(2.19 \pm 0.22 \pm 0.32) \times 10^{-7}$\,cm$^{-2}$\,s$^{-1}$, which is about one third of the flux of the EGRET source 3EG~J1048$-$5840. Figure \ref{fig:sed_j1048} shows both the overall fit between 0.1 and 20 GeV (solid line), and the spectral points obtained for 6 logarithmically-spaced energy bins and performing spectral analysis in each interval, assuming a power law shape for the source. To check the assumption of a cut-off energy in the spectrum, we also modeled the pulsar with a simple power law of the form $dF/dE = N_{0} E^{-\Gamma}$. The chance probability to incorrectly reject the hypothesis of a pure power law spectrum is $\sim$ 10 $\sigma$.

The pulsar is listed in the Bright Source List of the \textit{Fermi} LAT \citep{abdo09d} as 0FGL~J1047.6$-$5834, which is located at (RA,Dec) = (161.922\degr, $-$58.577\degr) with a 95\% confidence level radius of 0.138$^{\circ}$. There is another LAT point source located $\sim$1$^{\circ}$ away, 0FGL~J1045.6$-$5937. We modeled this unidentified source with a simple power law. The best fit result gives a spectral index $2.2 \pm 0.1 \pm 0.1$ and an integral photon flux of $(4.49 \pm 0.40 \pm 0.80) \times 10^{-7}$\,cm$^{-2}$\,s$^{-1}$. The sum of the fluxes of this source plus PSR~J1048$-$5832 is $\sim$ 6.7 $\times$10$^{-7}$ cm$^{-2}$s$^{-1}$ which is consistent with the flux of 3EG~J1048$-$5840 of $(6.2 \pm 0.7) \times 10^{-7}$\,cm$^{-2}$\,s$^{-1}$. The EGRET source 3EG~J1048$-$5840 as well as the COS--B source 2CG 288--00 \citep{swan81} were apparently made up of these two sources, which have now been resolved by the \textit{Fermi} LAT.

\subsubsection{Spectrum of PSR~J2229+6449}

\begin{figure*}
\begin{center}
\epsscale{1.}
\plotone{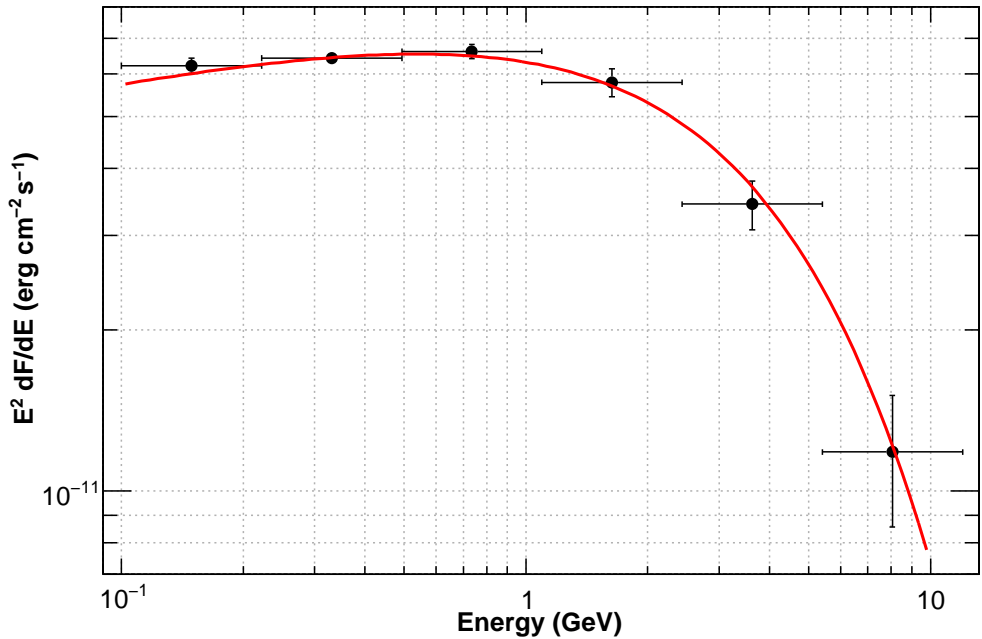}
\caption{Spectral energy distribution for PSR~J2229+6114 as fit by ``gtlike''assuming a power-law with an exponential cutoff (solid line). The spectral points were obtained for 6 logarithmically-spaced energy bins and performing spectral analysis in each interval, assuming a power law shape for the source. The errors bars are statistical only.}
\label{fig:sed_j2229}
\end{center}
\end{figure*}

Initially, a 15 degree circular region around the source position and the same Galactic diffuse background used for PSR~J1048$-$5832 were modeled to fit PSR~J2229+6114. However, some structures around the object were not taken into account in the Galactic model, and overestimate the flux of the pulse regions. We finally adopted a model for the Galactic diffuse emission based on six galactocentric 'ring' maps of N(H~I) and W(CO) and on the spatial distribution of the  inverse Compton intensity modeled by GALPROP. This intensity as well as the gamma-ray emissivities per ring were adjusted to maximize agreement with the observations, taking into account detected point sources of gamma rays.  This approach is similar to that described by \citet{casandjian08} for modeling EGRET data, and is being used by the LAT team to fit the model of Galactic diffuse emission for the first public release.\\
PSR~J2229+6114, referenced as the LAT point source 0FGL~J2229.0+6114 in the Bright Source List, was modeled by a power law with a simple exponential cut-off (Eq.~\ref{eq:spectrum}), between 0.15 and 0.65 in phase. Figure~\ref{fig:sed_j2229} (solid line) shows the phase-averaged spectral energy distribution from the likelihood fit, with $N_{0}$ = $(5.2 \pm 0.4 \pm 0.1) \times 10^{-8}$\,cm$^{-2}$\,s$^{-1}$\,GeV$^{-1}$, a spectral index $\Gamma$ = $1.85 \pm 0.06 \pm 0.10$ and a cut-off energy $E_{c} = 3.6^{+0.9}_{-0.6} \pm 0.6 $\,GeV. From this, we estimated an integral photon flux of $(3.77 \pm 0.22 \pm 0.44) \times 10^{-7}$\,cm$^{-2}$\,s$^{-1}$. This value is 10\% lower than the flux of $(4.13 \pm 0.61) \times 10^{-7}$\,cm$^{-2}$\,s$^{-1}$ of the EGRET source 3EG~J2227+6122 obtained for a power law with a spectral index of $2.24 \pm 0.14$ and larger than the flux of $(2.6 \pm 0.4) \times 10^{-7}$\,cm$^{-2}$\,s$^{-1}$ measured by AGILE \citep{pellizzoni09}. We also fit the LAT data to a simple power-law model, yielding a flux of $(4.54 \pm 0.14) \times 10^{-7}$\,cm$^{-2}$\,s$^{-1}$ and a spectral index of $2.25 \pm 0.02$. Note that the spectral model using an exponential cut-off is better constrained with a difference between the log likelihoods of $\sim 9\sigma$, rejecting the power-law hypothesis. For both PSR~J1048--5832 and PSR~J2229+6114 the statistics available is not enough to significantly rule out a super-exponential cutoff in favor of a simple exponential cutoff.

As a first search for unpulsed emission from the Boomerang nebula, we fitted a point-source to the off-pulse data at the radio pulsar position in the energy band 0.2\ --\ 100 GeV. No signal was observed from the PWN. After scaling to the full pulse phase, we derived a 95\% confidence level upper limit on the flux of $4.0\times 10^{-8}$\,cm$^{-2}$\,s$^{-1}$.

\section{Discussion}

The $\gamma$-ray light curves of PSR~J1048$-$5832 and PSR~J2229+6114 cover a wide range in phase, suggesting that the $\gamma$-ray beams cover a large solid angle. This in turn seems to favor the outer magnetospheric emission models, in particular the outer gap (OG, \citealt{romani96}) and the slot gap (SG, \citealt{muslimov04}) models.

The $\gamma$-ray luminosity is given by 
\begin{equation}
L_{\gamma} = 4\pi f_{\Omega}(\alpha,\zeta_{E}) F_{E,obs} d^{2}, 
\end{equation}
where $F_{E,obs}$ is the observed energy flux at the Earth line of sight (at angle $\zeta_E$ to the rotation axis), $d$ is the pulsar distance, and $f_{\Omega}(\alpha,\zeta_{E})$ is the beaming correction factor that depends on the geometry of the emission pattern. The factor $f_{\Omega}(\alpha,\zeta_{E})$ is a function of the pulsar magnetic inclination $\alpha$, and is model sensitive. It is given by \citep{watters09}:
\begin{equation}\label{eq:fbeam}
f_{\Omega}(\alpha,\zeta_{E}) = \frac{\int F_{\gamma}(\alpha;\zeta,\phi)\sin(\zeta)d\zeta d\phi}{2 \int F_{\gamma}(\alpha;\zeta_{E},\phi) d\phi}
\end{equation}
where $F_{\gamma}(\alpha;\zeta,\phi)$ is the radiated flux as a function of the viewing angle $\zeta$ and the pulsar phase $\phi$. In the ratio $f_{\Omega}$, the numerator is the total emission over the full sky, and the denominator is the expected phase-averaged flux for the light curve seen from Earth. For the polar cap model \citep{daugherty96}, the $\gamma$-ray emission originates at a few stellar radii from the surface, implying an emission with a small solid angle, that is, $f_{\Omega} \ll$ 1. For both OG and SG models, where the emission is far away from the neutron star, the resulting $f_{\Omega}$ values can be near unity, and even exceed 1 when the dominant sampling of the flux $F_{\gamma}$ arises for viewing angles $\zeta$ quite disparate from $\zeta_E$.


For PSR~J1048$-$5832, the observed integral energy flux obtained by integrating Equation \ref{eq:spectrum} times the energy is  $(19.4 \pm 1.0 \pm 3.1) \times 10^{-11}$\,erg\,cm$^{-2}$\,s$^{-1}$, leading to a $\gamma$-ray luminosity of $2.1 \times 10^{35}$ $f_{\Omega}$ ($d$/3~kpc)$^{2}$ and an efficiency $\eta$ = $L_{\gamma}/\dot{E}$ = 0.10 $f_{\Omega}$ ($d$/3~kpc)$^{2}$. 

The peak separation in the $\gamma$-ray light curve and the $\gamma$-ray efficiency are useful to place constraints in the now-favored outer magnetospheric emission models. Using the $\gamma$-ray light curve ``Atlas'' of \citet{watters09} we can estimate an allowed range $\alpha \sim$ 60\degr -- 85\degr~, $\zeta \sim$ 70\degr -- 80\degr~, and $f_{\Omega} \sim$ 0.7 -- 1.1 for the OG model, and $\alpha \sim$ 50\degr -- 75\degr~, $\zeta \sim$ 50\degr -- 75\degr~, and $f_{\Omega} \sim$ 1.1 for the SG (Two-Pole Caustic) model. The variation of the radio polarization position angle for a pulsar constrains the impact angle $\beta_{E}$ = $\alpha$ - $\zeta_{E}$. Fitting the rotating vector model \citep{radhna69} to the PSR J1048$-$5832 radio data, a value of $\beta_{E}$ smaller than 10\degr~has been derived \citep{kara05,welte08a}. Both theoretical models thus have a good range of possible solutions, but these estimates assume the efficiency relation $\eta \simeq (10^{33}/\dot{E})^{0.5}$ of \citet{watters09}, which gives $\eta \simeq$ 0.02, about a factor of five smaller than that derived for this pulsar from its inferred luminosity. Assuming our measured quantity $\eta$ = 0.10 the allowed range remains about the same for the Two-Pole Caustic (TPC) model and the range of $\alpha$ shifts to $\sim$ 70\degr\ --\ 90\degr~ for the OG model. The phase lag between the radio pulse and the first $\gamma$-ray pulse is in agreement with both OG and TPC or SG models. \\

For PSR~J2229+6114, the observed integrated energy flux is $(23.7 \pm 0.7 \pm 2.5) \times 10^{-11}$\,erg\,cm$^{-2}$\,s$^{-1}$. Thus, we estimate a $\gamma$-ray luminosity $L_{\gamma}$ = 2.6$\times$10$^{35}f_{\Omega}$($d$/3~kpc)$^{2}$ erg s$^{-1}$, and deduce an efficiency $\eta_\gamma$ = 0.011 $f_{\Omega}$ ($d$/3~kpc)$^{2}$, which is a factor 1.6 larger than the efficiency estimated by the relation $\eta \simeq (10^{33}/\dot{E})^{0.5}$. Note that with the estimated SNR distance of 0.8\,kpc the efficiency decreases to 0.001 $f_{\Omega}$, while with the distance derived from the DM (7.5\,kpc) the efficiency increases to 0.07 $f_{\Omega}$, emphasizing the importance of the distance determination.

As for PSR J1048$-$5832, we can compare light curves with the geometrical models of \citet{watters09} to derive constraints on the geometry. For outer magnetosphere models with $\eta_{\gamma} \sim 0.01$,~both TPC and OG geometries can deliver single $\gamma$-ray pulses at a range of angles. Single pulse solutions appear for $\alpha \sim 20^{\circ} - 55^{\circ}$, $\zeta \sim 25^{\circ} - 50^{\circ}$ for the TPC/SG geometry and over the range $\alpha \sim 45^{\circ} - 80^{\circ}$ , $\zeta \sim 35^{\circ} - 70^{\circ}$ and $f_{\Omega} \sim$ 0.47 - 1.08 for the OG model. For PSR J2229+6114, we have an additional geometrical constraint from modeling of the X-ray PWN torus surrounding the pulsar, as measured by {\em Chandra}; \citet{NgRomani04,NgRomani08} fit these data to determine a viewing angle $\zeta = 46\degr \pm 2\degr \pm 6\degr$. This is in the overlap range consistent with both the TPC and the OG single pulse solutions. Finally, although there is some uncertainty in determining the precise phase of the magnetic axis from the radio pulse, the lag of the strong $\gamma$-ray peak from the radio peak ($\phi \approx 0.50$) provides additional information. Examining the sample light curves in Figures 9 \& 10 of \citet{watters09} we find that single peaks at this phase are found only when they can be identified as the normal `P2' component, with a `bridge' of emission to earlier phases and `P1' faint or missing. Such light curves appear only for small efficiency $\eta < 0.03$ and only for a modest range of angles: $\alpha \sim 50^{\circ} - 70^{\circ}$, $\zeta \sim 40^{\circ} - 50^{\circ}$ (TPC/SG, with P1 faint) or $\alpha>45^{\circ}$, $\zeta = 45^{\circ} - 50^{\circ}$ (OG, with P1 absent). This is precisely the $\zeta$ range required by the X-ray torus fitting; including the small $\beta_{E}$ constraint from the detection of radio emission, we see that the radio, X-ray and $\gamma$-ray data are all consistent with $\alpha = 55^{\circ} \pm 5^{\circ}$, $\zeta = 45 \pm 5^{\circ}$, implying $f_{\Omega} \approx 1$ for both models. Thus low $\gamma$-ray efficiency, high altitude emission and a well constrained viewing geometry together provide a consistent picture for gamma-ray light curve shape. On the other hand, the gamma-ray light curve does not seem to match the the clearly double-peaked X-ray light curve. However, we have argued that the gamma-ray light curve is intrinsically double-peaked, with the first peak missing or very weak, and the appearance of a possible first peak in the $>$ 100 MeV and 0.1 - 0.3 gamma-ray light curves (signal/$\sqrt{{\rm background}} =$ 2.5\,$\sigma$) that is around the phase of the  first X-ray peak is intriguing. But the second gamma-ray peak is not in phase with the second X-ray peak. As the radiation mechanisms contributing to the emission in X-ray and LAT wavebands may differ, and at a given magnetospheric colatitude, the anisotropy of each emission mechanism is both energy-dependent and altitude-dependent, one does not expect the pulse profiles to be achromatic.  Therefore, more detailed physical radiation models are required to understand the energy dependence of the LAT light curve and to explain the lower energy X-ray light curve. \\

This evidence against low altitude emission in these pulsars can also be supplemented by constraints of a separate physical origin. The observation of photons out to beyond 5 GeV precludes any dominant action of magnetic pair creation $\gamma\to e^+e^-$ in the emission region at energies below this. Accordingly, the maximum observed energy, $\epsilon_{\rm max}$GeV, provides a lower bound to the altitudes of emission, since it must lie below any threshold energy for a super-exponential $\gamma$-B pair production turnover.  Such constraints have been obtained for $Fermi$ detections of the Vela pulsar \citep{abdo09a}, and PSR J1028-5819 \citep{abdo09e}, indicating that the super-GeV emission must originate above at least 2.2 and 2.1 stellar radii, respectively. Using a standard polar cap model estimate for the minimum emission height of $r\gtrsim (\epsilon_{\rm max} B_{12}/1.76\hbox{GeV})^{2/7}\, P^{-1/7}\, R_{\ast}$ (e.g. inverting Eq. [1] of \citet{baring04}), for a surface polar field strength of $10^{12}B_{12}$G, the PSR J1048-5832 spin-down parameters ($P=0.1237$\,s, $B_{12}=3.5$) together with the maximum observed energy $\epsilon_{\rm max}\sim 9$\,GeV in Figure~3 yield $r\gtrsim 3.1 R_{\ast}$. This drops to $r > 2.1 R_{\ast}$ if $E_c\sim 2.3$\,GeV is deployed when defining this bound (as was the case for the aforementioned \textit{Fermi} pulsar detections). Similarly, for PSR J2229+6114 with $P=0.0516$\,s, $B_{12}=6.4$ and the maximum observed energy $\epsilon_{\rm max}\sim 8$\,GeV in Figure~4, the altitude constraint is $r\gtrsim 4.0 R_{\ast}$. For either case, clearly these bounds preclude emission very near the stellar surface, adding to the advocacy for a slot gap or outer gap acceleration locale for the emission in both pulsars.

\section{Conclusion}
Although PSR~J1048$-$5832 and PSR~J2229+6114 are both Vela-like in age and spin characteristics, their light curves and derived emission geometries are quite different. Table \ref{tab:sumres} summarises the main quantities measured for both pulsars. The double-peaked light curve of PSR J1048$-$5832 is nearly identical to that of Vela, whereas its derived efficiency is a factor of 10 larger than that of Vela \citep{abdo09a}, adopting $f_{\Omega} = 1$ and $d = 3$\,kpc. On the contrary, the $\gamma$-ray efficiency of J2229+6114 is very similar to that of the Vela pulsar, but the pulsar shows a single, large peak similar to PSR~J0357+32 discovered by searching for pulsations at the positions of bright $\gamma$-ray sources \citep{abdo09d}. Note that the efficiency of PSR~J2229+6114 would be smaller at the distance of about 1\,kpc that some authors suggest.

\begin{table*}[t]
\caption{This table summarises the results of the timing and spectral analysis of PSR~J1048$-$5832 and PSR J2229+6114. Statistical and systematics errors are reported.}
\vspace{0.1cm}
\begin{center}

\begin{tabular}{llcc}
\hline
Analysis & Parameters & PSR~J1048$-$5832 & PSR~J2229+6449  \\
\hline
\hline       
\textbf{Timing results} & Number of pulsed $\gamma$ rays & 933 $\pm$ 93 & 1365 $\pm$ 97\\
       & Peak position ($\phi$)& 0.15 $\pm$ 0.01 $\pm$ 0.0001 (P1) & 0.49 $\pm$ 0.01 $\pm$ 0.001 \\
       &              & 0.57 $\pm$ 0.01 $\pm$ 0.0001 (P2) & \\
       & Peak separation ($\Delta$) & 0.42 $\pm$ 0.01 $\pm$ 0.0001 & \\
       & Peak FWHM& 0.06 $\pm$ 0.01 (P1) & 0.23 $\pm$ 0.03 \\
       &          & 0.10 $\pm$ 0.02 (P2) &  \\
\hline
       \textbf{Spectral results}
       & $^{a}$F (10$^{-7}$ cm$^{-2}$s$^{-1}$) & 2.19 $\pm$ 0.22 $\pm$ 0.32 & 3.77 $\pm$ 0.22 $\pm$ 0.44 \\
       & $^{b}$F$_{E}$ (10$^{-11}$ erg cm$^{-2}$s$^{-1}$) & 19.4 $\pm$ 1.0 $\pm$ 3.1 & 23.7 $\pm$ 0.7 $\pm$ 2.5 \\
       & $\Gamma$& 1.38 $\pm$ 0.06 $\pm$ 0.12 & 1.85 $\pm$ 0.06 $\pm$ 0.10 \\
       & $^{c}$E$_{c}$ (GeV)& 2.3$^{+0.3}_{-0.4}$ $\pm$ 0.3 & 3.6$^{+0.9}_{-0.6}$ $\pm$ 0.6 \\
       & L$_{\gamma}$ (10$^{35}$ erg s$^{-1}$) & 2.1 $f_{\Omega}$ (d/3kpc)$^{2}$ & 2.6 $f_{\Omega}$ (d/3kpc)$^{2}$ \\
       & $\eta_{\gamma}$ & 0.10 $f_{\Omega}$ (d/3kpc)$^{2}$& 0.011 $f_{\Omega}$ (d/3kpc)$^{2}$ \\
\hline
\end{tabular}
\tablenotetext{a}{Integral photon flux ($E>$0.1 GeV)}
\tablenotetext{b}{Integral energy flux ($E>$0.1 GeV)}
\tablenotetext{c}{Energy of an exponential cut-off to a power-law spectrum with index $\Gamma$.}
\end{center}
\label{tab:sumres}
\end{table*}

With the growing number of detected $\gamma$-ray pulsars, we are beginning to sample a wider variety of emission and viewing geometries, and pulsar ages. The range of light curve morphologies should allow improved constraints on high-energy emission models and a better understanding of the pulsar magnetospheric structure and acceleration process. For example, while many young pulsars, like J1048$-$5832, show Vela-type light curves, a small number are similar to J2229+6114, with a strong P2 component and a weak or absent P1 \citep{welte09b}. Mapping the angle range over which P1 is missing, especially when viewing angle constraints are available, can help narrow down the high altitude emission zone. A larger pulsar sample also allows a study of evolution of the $\gamma$-ray beaming and efficiency with pulsar age:~the pulsars seen with \textit{Fermi}, not including millisecond pulsars, span ages from 10$^{3}$ to $2 \times 10^{6}$\,yr \citep{abdo09i}, with hopes to extend that to larger $\tau$ (lower $\dot{E}$) as observations continue. Analysis of the population of pulsars with interpulses \citep{welte08b} and radio polarization data \citep{tauris98} have given hints that the magnetic inclination is larger for young pulsars and decreases with age. Thus we may even probe evolution of magnetic alignment during pulsar spindown. 


\acknowledgments

The \textit{Fermi} LAT Collaboration acknowledges generous ongoing support from a number of agencies and institutes that have supported both the development and the operation of the LAT as well as scientific data analysis. These include the National Aeronautics and Space Administration and the Department of Energy in the United States, the Commissariat \`a l'Energie Atomique and the Centre National de la Recherche Scientifique / Institut National de Physique Nucl\'eaire et de Physique des Particules in France, the Agenzia Spaziale Italiana and the Istituto Nazionale di Fisica Nucleare in Italy, the Ministry of Education, Culture, Sports, Science and Technology (MEXT), High Energy Accelerator Research Organization (KEK) and Japan Aerospace Exploration Agency (JAXA) in Japan, and the K.~A.~Wallenberg Foundation, the Swedish Research Council and the Swedish National Space Board in Sweden.

Additional support for science analysis during the operations phase is gratefully acknowledged from the Istituto Nazionale di Astrofisica in Italy.

The Parkes radio telescope is part of the Australia Telescope which is funded by the Commonwealth Government for operation as a National Facility managed by CSIRO. We thank our colleagues for their assistance with the radio timing observations.

The Green Bank Telescope is operated by the National Radio Astronomy Observatory, a facility of the National Science Foundation operated under cooperative agreement by Associated Universities, Inc.

The Lovell Telescope is owned and operated by the University of Manchester as part of the Jodrell Bank Centre for Astrophysics with support from the Science and Technology Facilities Council of the United Kingdom.



\end{document}